\documentclass[letterpaper]{article} 
\usepackage{aaai24}  
\usepackage{times}  
\usepackage{helvet}  
\usepackage{courier}  
\usepackage[hyphens]{url}  
\usepackage{graphicx} 
\usepackage{natbib}  
\usepackage{caption} 
\usepackage{algorithm}
\usepackage{algorithmic}

\usepackage{newfloat}
\usepackage{listings}
\DeclareCaptionStyle{ruled}{labelfont=normalfont,labelsep=colon,strut=off} 
\floatstyle{ruled}
\newfloat{listing}{tb}{lst}{}
\floatname{listing}{Listing}
\usepackage{color} 
\nocopyright 

\newcommand\blfootnote[1]{%
  \begingroup
  \renewcommand\thefootnote{}\footnote{#1}%
  \addtocounter{footnote}{-1}%
  \endgroup
}

\title{How COVID-19 has Impacted the Anti-Vaccine Discourse:\\ A Large-Scale Twitter Study Spanning Pre-COVID and Post-COVID Era}

\author{
   Soham Poddar\textsuperscript{\rm $\S$},
   Rajdeep Mukherjee\equalcontrib,
   Subhendu Khatuya\equalcontrib,
   Niloy Ganguly,
   Saptarshi Ghosh
}
\affiliations{
   Indian Institute of Technology Kharagpur, West Bengal 721302, India.\\
   
   \textsuperscript{\rm $\S$} sohampoddar@kgpian.iitkgp.ac.in   
}

\usepackage{multirow}
\usepackage{amsmath}
\usepackage{amssymb}
\usepackage{dsfont}
\usepackage{subcaption}
\usepackage{tikz}
\usepackage{soul}
\usepackage{adjustbox}
\usepackage{booktabs}
\usepackage{makecell}

\newcommand{\new}[1]{#1}

\newcommand{\classifier}[0]{CoV-Ent}
\newcommand{\genclassifier}[0]{CoV-Gen}
\newcommand{\dataset}[0]{CaV-N}

\begin{document}

\maketitle
\blfootnote{\textbf{\textit{This work has been accepted at the 18th International AAAI Conference on Web and Social Media (ICWSM), 2024.}}}

\begin{abstract}

The debate around vaccines has been going on for decades, but the COVID-19 pandemic showed how crucial it is to understand and mitigate anti-vaccine sentiments. While the pandemic may be over, it is still important to understand how the pandemic affected the anti-vaccine discourse, and whether the arguments against \textit{non-COVID vaccines} (e.g., Flu, MMR, IPV, HPV vaccines) have also changed due to the pandemic. This study attempts to answer these questions through a large-scale study of anti-vaccine posts on Twitter. 
Almost all prior works that utilized social media to understand anti-vaccine opinions considered only the three broad stances of Anti-Vax, Pro-Vax, and Neutral. 
There has not been any effort to identify the specific reasons/concerns behind the anti-vax sentiments (e.g., side-effects, conspiracy theories, political reasons) on social media at scale. 
In this work, we propose two novel methods for classifying tweets into 11 different anti-vax concerns -- a discriminative approach (entailment-based) and a generative approach (based on instruction tuning of LLMs) -- which outperform several strong baselines. 
We then apply this classifier on anti-vaccine tweets posted over a 5-year period (Jan 2018 - Jan 2023) to understand how the COVID-19 pandemic has impacted the anti-vaccine concerns among the masses. 
We find that the pandemic has made the anti-vaccine discourse far more complex than in the pre-COVID times, and increased the variety of concerns being voiced. 
Alarmingly, we find that concerns about COVID vaccines are now being projected onto the non-COVID vaccines, thus making more people hesitant in taking vaccines in the post-COVID era.

\end{abstract}

\section{Introduction}
\label{sec:intro}

Vaccines are considered to be a crucial weapon for prevention of several deadly diseases. 
However, a fierce debate around vaccines has existed since the past several decades, with \textit{``Pro-Vaxxers''} supporting the use of vaccines, and \textit{``Anti-Vaxxers''} opposing vaccines. 
Both these groups actively share their opinions on vaccines over social media. With the onset of the COVID-19 pandemic, the Anti-Vaxxer movement gained momentum, and a lot more people started sharing their concerns about COVID vaccines during the years 2020-21. 
Though the COVID pandemic has subsided (WHO declared end to COVID-19 as a global health emergency on 5 May 2023\footnote{\url{https://news.un.org/en/story/2023/05/1136367}})
it is still important to understand whether the anti-vaccine discourse has changed as a result of the pandemic. 
This understanding is particularly important for several \textit{non-COVID vaccines} that still needs to be administered regularly for the well-being of the society. 
Many such non-COVID vaccines have been actively opposed since pre-COVID times, such as the MMR (Measles-Mumps-Rubella) vaccine, the IPV vaccine, the Flu vaccine, and so on~\cite{kata2012anti}. 
It is important to understand whether the anti-vaccine discourse around these non-COVID vaccines has also been affected by the pandemic, so that the evolving anti-vaccine concerns can be countered.

Many prior works have tried to understand the vaccine discourse through the lens of social media such as Twitter~\cite{gunaratne2019temporal,muller2019crowdbreaks,cotfas2021longest,poddar2022winds}. 
All these prior works have classified/labeled vaccine-related tweets based on their high-level stance towards vaccines -- Anti-Vax, Pro-Vax or Neutral (neither anti- nor pro-vax). 
However, there are several distinct fine-grained  concerns that people express towards vaccines. 
For instance, some people might be hesitant because of the potential \textit{side-effects} of vaccines, some may argue that the vaccines are \textit{not effective} enough, while some may refuse to take vaccines because of \textit{political reasons}.
Identifying these fine-grained anti-vaccine concerns is crucial to understanding the anti-vaccine discourse. 
However, very little work has been done on automating this task of detecting the specific concerns towards vaccines from textual posts (\new{details in the `Related Works' section}).
In fact, till recently, there was no dataset for developing ML/NLP models to identify specific anti-vaccine concerns from texts. 
\new{Our recent prior work~\cite{poddar2022caves} developed a dataset called `CAVES'} that labels tweets with 11 distinct anti-vaccine concerns (see Table~\ref{tab:classes}), which
has opened up the possibilities for exploring ML/NLP methods for detection of specific anti-vaccine concerns. 
In this work, we utilize this dataset to address the following Research Questions -- 
\textbf{(RQ1)~How can anti-vax social media posts (tweets) be accurately classified based on the specific anti-vaccine concern(s) voiced in them?} 
This multi-class, multi-label classification problem is particularly challenging since all the concerns fall under the broad umbrella of anti-vaccine content, and the vocabulary used in tweets voicing the different concerns is  similar and overlapping (as also pointed out in~\cite{poddar2022caves}).
\new{The very skewed distribution of the classes also adds to the challenge, with the `side-effect' class being present in $38.4\%$ of the tweets while the `conspiracy' and `ingredients' classes being present in $<5\%$ of the tweets.}

In this work, we propose novel methods for this classification problem. 
To distinguish effectively among the different classes/labels, we leverage the \textit{label semantics}, i.e., the descriptions of the labels used by human annotators to label the tweets (the descriptions are stated in Table~\ref{tab:classes}).
Specifically, we employ two approaches -- (i)~a discriminative approach that formulates the classification task as an entailment task, and 
(ii)~a generative approach based on instruction tuning of a Large Language Model (LLM) -- both utilizing the label descriptions.
Our proposed classifiers outperform several strong baselines (that did not attempt to use label descriptions) by \textbf{over 23\%} in terms of \textbf{Macro-F1} score.

Subsequently, we use our developed classifier to address the following research questions (RQs) to gain a detailed understanding of the effects of the COVID-19 pandemic on the anti-vaccine discourse. We ask --
\textbf{(RQ2)~Did the anti-vaccine discourse (over social media) change as a result of the COVID-19 pandemic?} If so, how?  
After the pandemic has subsided, has the anti-vaccine discourse returned to its pre-COVID state, or are there still lingering differences? 
Furthermore, given the continued importance of non-COVID vaccines (e.g., MMR, Flu, IPV vaccines) in the future, we also ask -- \textbf{(RQ3)~Has the anti-vaccine discourse around \textit{non-COVID} vaccines also been impacted by the COVID-19 pandemic?} 
Note that, though there have been several studies on the social media discourse around COVID vaccines, to our knowledge, this is the first large-scale study on how the COVID-19 pandemic affected the discourse around non-COVID vaccines.

Our study uncovers several changes that have come in the anti-vaccine discourse since the pandemic. 
During pre-COVID times, the anti-vaccine discourse was predictable and mostly discussed the `side-effects' of the vaccines. But, \textit{since the pandemic, the concerns have become much more diversified} with several other concerns also being discussed prominently, such as the vaccines are `rushed', `ineffective' and influenced by `political' factors. 
In the post-COVID period, we see the trends slowly moving back towards those in the pre-COVID times, but there still remain prominent differences. 
Importantly, we find that the discourse around non-COVID vaccines has also been largely affected by the pandemic. For instance, in the post-COVID times, there is a lot more discussion on these vaccines being ineffective, whereas the concerns with these vaccines being mandatory has reduced as compared to the pre-COVID times.
In fact, we observe that the \textit{anti-vaccine discourse around non-COVID vaccines has become much more complex since the pandemic}. 
For many people, COVID has enhanced their prior concerns about non-COVID vaccines -- this is particularly true for the traditional anti-vaxxers who were hesitant about taking vaccines even during pre-COVID times.
Alarmingly, we find that several \textit{new concerns that arose about COVID vaccines are now being projected onto non-COVID vaccines} as well. For instance, the mRNA technology that was used in some COVID vaccines is now allegedly said to be used in Flu vaccines. 
As a result, we find \textit{a set of Twitter users (whom we call `converted anti-vaxxers') who supported vaccines in pre-COVID times, but are now unwilling to take vaccines any further}. 
We study the opinion changes of these two sets of users to answer the last research question \textbf{(RQ4)~What are the concerns of traditional and converted anti-vaxxers after the pandemic?}

\vspace{2mm}
\noindent \textbf{Contributions:}
Overall, our contributions in this paper are as follows:- 
(1)~We propose novel classifiers for labeling a anti-vax tweet with fine-grained anti-vax concerns, leveraging the label descriptions. Our classifiers outperform several strong baselines for this challenging multi-label classification task. We also show that our classifiers perform well on tweets related to both COVID vaccines and non-COVID vaccines.
(2)~We perform a long-term analysis of the anti-vaccine discourse, spanning pre-COVID, COVID and post-COVID times, to analyze how the anti-vax concerns of people have changed due to the pandemic. To our knowledge, this is the first study on how the COVID-19 pandemic affected the discourse around \textit{non-COVID vaccines} as well.
(3)~We identify several important changes in the anti-vaccine discourse since the pandemic, some of which are lingering even after the pandemic. Importantly, we uncover new concerns emerging about non-COVID vaccines (such as the Flu vaccine), which is causing more people to become hesitant towards these vaccines.
\new{
The datasets and codes for our work have been made available to promote further research.\footnote{\url{https://github.com/sohampoddar26/covid-impact-antivax}}
}

We believe this work provides an important step in identifying specific anti-vax concerns of netizens from their social media posts, and presents insights that are crucial in alleviating people's vaccine concerns in the future.

\section{Related Works}
\label{sec:litsurvey}

In this section, we briefly discuss some related works in the areas of vaccine-related information on social media and some classification methods.

\begin{table}[!t]
    \centering
    \small
    \begin{tabular}{|p{0.948\linewidth}|}
    \hline
    \textbf{Conspiracy:} Deeper Conspiracy -- The tweet suggests some deeper conspiracy, and not just that the Big Pharma want to make money (e.g., vaccines are being used to track people, COVID is a hoax). \\
    \hline
    
    \textbf{Country:} Country of origin -- The tweet is against some vaccine because of the country where it was developed/manufactured. \\
    \hline
    
    \textbf{Ineffective:} Vaccine is ineffective -- The tweet expresses concerns that the vaccines are not effective enough and are useless. \\
    \hline
    
    \textbf{Ingredients:} Vaccine Ingredients/technology -- The tweet expresses concerns about the ingredients present in the vaccines (eg. fetal cells, chemicals) or the technology used (e.g., mRNA vaccines can change your DNA). \\
    \hline
        
    \textbf{Mandatory:} Against mandatory vaccination -- The tweet suggests that vaccines should not be made mandatory. \\
    \hline
    
    \textbf{Pharma:} Against Big Pharma -- The tweet indicates that the Big Pharmaceutical companies are just trying to earn money, or is against such companies in general because of their history. \\
    \hline
     
    \textbf{Political:} Political side of vaccines -- The tweet expresses concerns that the governments/politicians are pushing their own agenda though the vaccines. \\
    \hline
    
    \textbf{Religious:} The tweet opposes vaccines due to religious reasons. \\
    \hline
    
    \textbf{Rushed:} Untested/Rushed Process -- The tweet expresses concerns that the vaccines have not been tested properly or that the published data is not accurate.\\
    \hline
    
    \textbf{Side-effect:} Side Effects/Deaths -- The tweet expresses concerns about the side effects of the vaccines, including deaths. \\
    \hline
    
    \textbf{Unnecessary:} The tweet indicates vaccines are unnecessary, or that alternate cures are better. \\
    \hline
    
    \textbf{None:} No specific reason stated in the tweet, or some reason other than the given ones. \\
    \hline
    \end{tabular} 
    \caption{The different classes/labels (concerns or objections towards vaccines) in the CAVES dataset~\cite{poddar2022caves} along with their descriptions.}
    \label{tab:classes}
\end{table}

\begin{table}[tb]
    \centering
    \small
    \begin{tabular}{|p{58.5mm}|p{16.6mm}|}
    \hline
    \textbf{Tweet Excerpt} & \textbf{Labels} \\
    \hline
    STOP TAKING \textit{TOXIC VAX} and {\it expose COVID hoax} and murders with morphine and ventillators. {\it there is No covid}! & ingredients, conspiracy, unnecessary\\ 
    \hline
    Please {\it don't push vaccine on us make it voluntary}. We don't trust anything to do with {\it Bill Gates} pushing their agenda of {\it vaccine chips}!! & pharma, mandatory, ingredients\\
    \hline
    The reason insurance companies won't pay out if you experience the inevitable {\it adverse reactions, including death} is because it is an {\it ``Experimental Vaccine''} & side-effect, rushed \\
    \hline
    Would you want the {\it Russian vaccine}? If not, you shouldn't want one that's been {\it pushed through for political reasons} either. & political, country\\
    \hline
    I'm NOT taking your damn vaccine. Keep your conspiracy out of my veins! & none \\
    \hline
    \end{tabular}
    \caption{Examples of tweets with their labels and explanations, from the CAVES dataset. The explanations for different labels are highlighted in \textit{italics}.} 
    \label{tab:multilabelexample}
\end{table}

\vspace{2mm}
\noindent \textbf{Vaccine studies on social media:}
Since the onset of the COVID-19 pandemic, 
social media discussions around vaccines have increased rapidly. 
In fact, at the start of the pandemic during 2020, there was a huge surge of anti-vaccine opinions ~\cite{burki2020online,bonnevie2020quantifying,durmaz2022dramatic} which started decreasing again as the COVID-19 vaccines rolled out in early 2021~\cite{poddar2022winds}.
While prior works have analyzed vaccine sentiments during 2020-21,
the lingering effects in the post-COVID era remain to be examined (which we do in this paper).

Furthermore, though many prior works have studied people's opinions on COVID vaccines (as stated above), to our knowledge, there are very few studies that try to understand how the anti-vaccine discourse around \textit{non-COVID vaccines} has evolved during and after the pandemic. 
A few very recent works talk about the impact of the COVID-19 pandemic on the traditional non-COVID vaccines, such as MMR, IPV, etc. However they only comprise of anectodal evidence~\cite{altman2023factors,knijff2023parental} and rudimentary survey-based analyses~\cite{rivera2023attitudes}.
Our study is the first to analyse the impact of COVID-19 on opinions towards the non-COVID vaccines on a large scale over social media.

\vspace{2mm}
\noindent \textbf{Classification of vaccine-related social media posts:}
Most prior works categorized vaccine-related posts into three broad classes -- Pro-Vax (supports vaccines), Anti-Vax (against vaccines) and Neutral~\cite{muller2019crowdbreaks,cotfas2021longest,poddar2022winds}, while a recent work~\cite{mu2023vaxxhesitancy} split the anti-vax class into ``strongly anti-vax'' and ``vax-hesitant''. 
However, prior works have shown that there exist several fine-grained anti-vaccine concerns~\cite{fasce2023taxonomy}; for instance, some people worry about side-effects of vaccines, while others are against the politicization of vaccines.
Thus, simply studying the broad classes does not give us a complete picture of the different shades of anti-vaccine content on social media.

A few prior works tried to understand these fine-grained reasons for vaccine hesitancy, mostly by manual analysis~\cite{bonnevie2020quantifying} and unsupervised topic modelling~\cite{praveen2021analyzing,poddar2022winds,hwang2022vaccine}.
\new{We recently developed the CAVES dataset~\cite{poddar2022caves} that enables training deep neural models to accurately identify vaccine concerns from tweets.} 
To our knowledge, no prior work has tried to apply a supervised classifier on social media posts to derive insights about changes in the anti-vaccine discourse.

\vspace{2mm}
\noindent \textbf{Multi-label Classification:}
Multi-label classification is a long-studied problem, and several approaches have been applied in various sub-domains of social media analysis, such as emotion detection from tweets~\cite{mukherjee2021understanding,ameer2023multi}, 
disaster mitigation~\cite{chowdhury2020cross} and symptom detection~\cite{jarynowski2021mild}. 
Looking out to general domains, entailment-based methods~\cite{wang2021entailment} and generative models~\cite{simig2022open} (which we apply in this work) have been applied to classification tasks.
However, the key novelty in our proposed methods -- incorporating label semantics into multi-label classification, has not been explored earlier in the domain of social media analysis.

\section{The CAVES Dataset for anti-vax concerns}

CAVES~\cite{poddar2022caves} consists of 9,921 anti-vaccine (anti-vax) tweets that convey hesitancy towards COVID-19 vaccines. 
Each tweet is labelled with specific concerns that the user (author of the tweet) expressed against the intake of COVID-19 vaccines (e.g., ineffectiveness, side effects, etc.). 
There are 12 different classes (labels), as detailed in Table~\ref{tab:classes}, with 11 of them representing specific anti-vax concerns, while the last one (`\textit{None}') signifying that no specific anti-vaccine concern is mentioned in the tweet.
For each label, there is an explanatory `description' (shown in Table~\ref{tab:classes}) that was given to the annotators while labeling the tweets during preparation of the dataset~\cite{poddar2022caves}.
Given that multiple anti-vax concerns may be mentioned in a tweet, about 20\% of the tweets in the dataset are associated with more than one label, whereas the remaining tweets have exactly one label.
We have reported a few examples of tweets along with their labels in Table~\ref{tab:multilabelexample}.
For each label associated with a tweet, the CAVES dataset also contains a separate explanation in the form of a phrase/span appearing in the tweet text.
Table~\ref{tab:multilabelexample} also shows the explanations (italicized parts) corresponding to each label associated with a tweet.

\vspace{2mm}
\noindent {\bf Train, validation, and test splits:} The CAVES dataset was already split (by iterative stratified sampling) into train (70\%), validation (10\%) and test (20\%) sets in our prior work~\cite{poddar2022caves}. We consistently use the same splits to train, validate and test all models in this work.

\section{Classifiers for anti-vax concerns}
\label{sec:methodology}

\begin{table*}[!t]
    \centering
    \small
    \begin{tabular}{|l|p{99mm}|l|l|}
        \hline
        \textbf{Premise (Tweet Text)} & \textbf{Hypothesis (Class Description)} & \textbf{Labels} &  \textbf{Target} \\
        \hline 
        \multirow{6}{27mm}{A vaccine that cant prevent reinfection? No thanks I trust my own immunity thats over 99.5\% effective.} & Vaccine is ineffective – The tweet expresses concerns that the vaccines are not effective enough and are useless. & ineffective & entailment \\
        \cline{2-4}
        & The tweet indicates vaccines are unnecessary, or that alternate cures are better. & unnecessary & entailment \\
        \cline{2-4} \cline{2-4}
        & Against mandatory vaccination – The tweet suggests that vaccines should not be made mandatory. & mandatory & contradiction \\
        \cline{2-4}
        & The tweet opposes vaccines due to religious reasons. & religious & contradiction \\
        \hline
    \end{tabular}
    \caption{Example of inputs and targets to train \classifier{}. `ineffective' and `unnecessary' are the ground truth labels for the given tweet, while `side-effect' and `mandatory' are randomly chosen wrong labels for contrastive/negative sampling. 
    }
    \label{tab:entail_IO}
\end{table*}

Most existing multi-label classifiers typically treat classes or labels as mere identifiers without taking into account any inherent label semantics. 
We adopted such conventional models like Roberta, BERT, etc. as our baseline approaches (detailed later in this section). 
However, the classes in our dataset closely relate to each other, as they all fall under the umbrella of `anti-vaccine' concerns. 
Consequently, tweets across different classes share significant semantic similarities, making it difficult for traditional multi-class classifiers to distinguish between them effectively. 
\new{For example, training a generative classifier (with standard greedy decoding) with only the labels as targets produced poor results, with the model not being able to accurately distinguish the classes.
}

We observe that the \textit{label descriptions} (shown in Table~\ref{tab:classes}) contain valuable information that can aid in better distinction among the classes.
The annotators also relied on this information to accurately assign labels while creating the dataset~\cite{poddar2022caves}. 
This insight motivated us to incorporate the label descriptions into our classifier design.
We propose two distinct classifiers which 
leverage the label descriptions through two different paradigms: (1)~a discriminative entailment-based method and (2)~a generative method based on instruction-tuning of LLMs.


\subsection{Discriminative classifier}

This approach, that we call \textbf{\classifier{}}, converts the classification problem to that of a sentence-pair \textit{Entailment Task} where a BERT-based classifier is trained to predict if the label descriptions can be inferred/entailed from the tweet text.

\vspace{2mm}
\noindent \textbf{Model Description:}
The model expects two input texts -- a \textit{premise}, 
and a \textit{hypothesis} containing some assumption based on the premise. 
The model is trained to predict if the hypothesis can be \textit{entailed} (inferred) from the premise or if it \textit{contradicts} the premise. 
In our case, the tweet text serves as the premise, and the description of an anti-vax concern class label (see Table~\ref{tab:classes}) serves as the hypothesis. 
The input to the model is a concatenation of these, separated by the \texttt{[SEP]} token. 
The target is set to either $0$ (entailment) or $1$ (contradiction).
\textbf{\classifier{}} consists of a COVID-Twitter-BERT based encoder~\cite{muller2020covid} with a binary classification head (fully connected linear layer) on top to predict if the input class description entails or contradicts the input tweet text.

\vspace{2mm}
\noindent \textbf{Training and Testing strategy:}
Our model \textit{\classifier{}} is trained with both \textit{positive} as well as \textit{negative} sampled data points. 
For each tweet $text_i$ in the training set, our \textit{positive} samples are formed using the ground truth labels associated with $text_i$, with \textit{`entailment'} as the target. 
We then randomly select $\mathcal{N}$ other labels (that are \textit{not} part of ground truth for $text_i$) to form the \textit{contrastive/negative} samples with target \textit{`contradiction'}. 
We call this number of negatively sampled data points per tweet \textit{`Negative Sampling Rate'} (a hyper-parameter). 
Example inputs and targets for training the model are illustrated in Table~\ref{tab:entail_IO} for a given tweet.

During testing, for each tweet in the test set, we use our trained model to obtain the predictions against each of the 12 possible hypotheses (corresponding to the 12 anti-vax classes in Table~\ref{tab:classes}). 
Those classes are predicted to be associated with the tweet where the model predicts \textit{``entailment''} as the output.  
If the model predicts ``\textit{none}'' along with other classes for some input, then we ignore the ``\textit{none}'' label. 


\subsection{Generative classifier}
\label{sub:method}

\begin{table*}[t]
\centering
\small
    \begin{tabular}{|p{24mm}|p{15.5mm}||l|l|}
        \hline
        \textbf{Tweet Text}  &\textbf{GT Labels} & \textbf{Input} & \textbf{Target}\\
        \hline 
        A vaccine that cant prevent reinfection? No thanks I trust my own immunity thats over 99.5\% effective. 
        &
        ineffective, unnecessary 
        &
        \parbox[t]{72mm}{
        Instruction: First read the task description. There could be multiple categories for a tweet. \\
        Task: Multi-label Text Classification \\
        Description: Generate label description for the given text. \\
        A vaccine that cant prevent reinfection? No thanks I trust my own immunity thats over 99.5\% effective. 
        }
        &
        \parbox[t]{48mm}{
        Vaccine is ineffective -- The tweet expresses concerns that the vaccines are not effective enough and are useless. \\
        The tweet indicates vaccines are unnecessary, or that alternate cures are better. 
        } \\
    \hline
    \end{tabular}
    \caption{Example of input (prompt and tweet-text) and target (the label descriptions, separated by full-stop) to train \genclassifier{}.}
    \label{table:instruction_example}
\end{table*}

\begin{table*}[t]
\centering
\small

       \begin{tabular}{|p{35mm}|p{42mm}|l|p{16mm}|}
       \hline
        \textbf{Tweet Text}  &\textbf{Generated Description} & \textbf{Matched Label Descriptions} & \textbf{Pred labels}\\
        \hline \hline
        A vaccine that cant prevent reinfection? No thanks I trust my own immunity thats over 99.5\% effective. &
        tweet indicates vaccines are unnecessary, or that alternate cures are better. Vaccine is ineffective -- The tweet expresses concerns & 
        \parbox[t]{66mm}{
        The tweet indicates vaccines are unnecessary, or that alternate cures are better. \\ Vaccine is ineffective -- The tweet expresses concerns that the vaccines are not effective enough and are useless.} &  
        unnecessary, ineffective \\
    \hline
    \end{tabular}
    \caption{Example of a text input and the corresponding generated output by \genclassifier{}.
    The generated text ($2^{nd}$ column) contains two descriptions separated by full-stop, which are matched to actual label descriptions leading to the predicted labels. 
    }
    \label{table:prediction_example}
\end{table*}

Our proposed generative framework for the multi-label classification task, that we call \textbf{\genclassifier{}},  is divided  into two stages: supervised generative phase, and an unsupervised description-to-label matching phase.

\vspace{2mm}
\noindent \textbf{Generating label descriptions:}
In the first stage, we formulate the problem as a generative task using LLMs, where given a tweet, the task is to accurately generate the descriptions of labels (and not the labels) associated with the tweet. 
Let $T_i = (t_i^{1},t_i^{2},t_i^{3},..., t_i^{n})$ be the $i^{th}$ tweet containing $n$ tokens, and let the tweet be labelled with $k$ labels ($c_1, c_2, ..c_k$). 
For each label $c_i$, we have a pre-defined label description as given in Table ~\ref{tab:classes}. 
In order to remove hashtags, mentions, emojis, URLs, etc, we pre-process the tweet using \textit{tweet-preprocessor}\footnote{https://pypi.org/project/tweet-preprocessor/} and obtain the filtered tweet $T_i^f$. 
We prepend an instruction prompt $IP$ containing a natural language description of the task, to the processed tweet as shown in the \textit{input} column of Table~\ref{table:instruction_example}. 
The modified input therefore takes the shape $IP || T_i^f$, where $||$ is a text concatenation operation. 
To construct our target, we concatenate the label descriptions, separating them with a \textit{full stop}.

We start with FLAN-T5 \cite{longpre2023flan,chung2022scaling} as the base LLM for this step. 
Our choice is driven by the fact that FLAN-T5 models are pre-trained using instruction tuning for a wide range of over 1.8K tasks. 
This extensive pre-training not only helps the models to align better to instructions for new downstream tasks, but also significantly reduces the number of fine-tuning steps required.
Moreover, FLAN-T5 models exhibit robust zero-shot and few-shot performance when compared to their non-instruction-tuned counterpart, T5 \cite{t5}. 
The model is trained to condition itself on the modified input $IP || T_i^f$, and generate the sequence of target descriptions token by token by means of auto-regressive decoding. 
Our training objective in this process is to minimize the cross-entropy loss between the generated tokens and the true tokens. 
The fine-tuning is conducted in a parameter-efficient way using Low-Rank Adaptation (LoRA)~\cite{hu2021lora}. 
LoRA freezes the pretrained model weights while introducing trainable rank decomposition matrices into each layer of the Transformer architecture. 
Effectively, it substantially reduces the number of parameters to be updated during the fine-tuning process, leading to significant reduction in training cost.

\vspace{2mm}
\noindent \textbf{Matching generated descriptions:}
In the second stage, we begin by breaking down the generated sequence of label descriptions into individual sentences, separated by full stops if present. 
Given that the generated description, corresponding to each such sentence, may not exactly match the ground truth label descriptions, we rely on a state-of-the-art pre-trained Sentence-BERT encoder~\cite{reimers2019sentence} to arrive at the final predicted label descriptions.
Specifically, we utilize this encoder to derive embeddings for each of the generated descriptions (sentences), denoted as $gendec_i^{f}$. 
Similarly, we also extract embeddings for the ground truth descriptions corresponding to all available 12 labels. 
The label description that exhibits the highest cosine similarity with the embedding of $gendec_i^{f}$ is considered to be the final predicted description. 
This approach takes into account the possibility of multiple label descriptions being generated by the model, and ensures that the most accurate label descriptions are finally assigned to the input tweet.
The corresponding labels can be easily obtained given the existence of a 1:1 correspondence between the labels and their descriptions.
Please refer to Table~\ref{table:prediction_example}, which presents an example of a tweet, its generated description, and the final predicted labels.


\subsection{Experiments and Results}
\label{sub:experiments}


\noindent \textbf{Baselines}:
We used the same baselines as in \citet{poddar2022caves}.
First, we tried some Transformer encoder models paired with a classification head on top for multi-label classification -- 
(1)~the \textbf{CT-BERT}~\cite{muller2020covid} model, and the (2)~\textbf{RoBERTa-Large}~\cite{liu2019roberta} model. 
Next, we incorporated the explanations (that are part of the CAVES dataset) as inputs to a model to see if it improves performance. 
Such models include a modified version of the (3)~\textbf{HateXplain}~\cite{mathew2021hatexplain} model, 
(4)~the \textbf{`\textit{Multi-task}'} model~\cite{poddar2022caves} and the (5)~\textbf{modified ExPred}~\cite{zhang2021explain} model which contain a shared CT-BERT encoder and multiple token and sequence classification layers that separately predict the labels and generate the explanations for each of the classes.

As another baseline, we used (6)~\textbf{GPT-3.5-Turbo (ChatGPT)} in a zero-shot setting -- we provided all class labels and descriptions as part of the prompt, along with a tweet text, and asked ChatGPT to predict which classes apply to the given tweet. 
We also tried a variation where we included an example tweet along with each class description, but this variation performed slightly worse and is thus omitted.

\vspace{2mm}
\noindent \textbf{Experimental Setup:} 
For \classifier, we used the COVID-Twitter-BERT (CT-BERT)~\cite{muller2020covid} encoder, which is a BERT-Large model pre-trained on a large set of COVID-related tweets. We observe the optimum results on the validation set for the negative sampling rate, $\mathcal{N} = 7$; hence this is what we use. 
For \genclassifier, we utilized the pre-trained checkpoints of FLAN-T5-Base from the \textit{Huggingface} Library~\cite{wolf-etal-2020-transformers}. For training this model with LoRA~\cite{hu2021lora}, the \textit{rank} for the trainable decomposition matrices was set to $2$. 
FLAN-T5-Base was trained with LoRA on the training set for only 5 epochs with a batch size of 8 and learning rate of $5e-4$. For all our experiments, the maximum input length was set to 128 and the maximum length of output to be generated was set to 50.
These hyperparameters were selected based on the best Macro-F1 results on the validation set of the CAVES dataset. 
All our experiments are performed on a Tesla V100 32GB GPU.

\vspace{2mm}
\noindent \textbf{Metrics:}
We used three standard metrics to evaluate the models. 
Given the set of predicted and gold standard labels for a tweet, we calculate the F1-score for each of the $12$ classes (as described in Table~\ref{tab:classes}) separately and find the (i)~Macro-average F1-score, and (ii)~Weighted-average F1-score with weights proportional to the class frequencies.
We also calculate the (iii)~Jaccard similarity between a tweet's predicted and gold standard label-sets and average it over all tweets in the test set.
All these standard metrics were calculated using the Scikit-Learn package~\cite{scikit-learn}.


\begin{table}[tb]
    \centering
    \small
    \begin{tabular}{|p{25mm}|c|c|c|}
    \hline
        \textbf{Model} & \textbf{Macro-F1} & \textbf{Weighted-F1} & \textbf{Jaccard} \\
        \hline \hline
        \multicolumn{4}{|l|}{\textit{Baselines that do not leverage label descriptions}}\\
        \hline
        RoBERTa-Large & 0.6626 & 0.7319 & 0.6949 \\
        CT-BERT & 0.6924 & 0.7419 & 0.7040 \\
        HateXplain & 0.6709 & 0.7245 & 0.6909 \\
        ExPred     & 0.6558 & 0.7105 & 0.6576 \\
        Multi-Task & 0.6823 & 0.7371 & 0.7018 \\
        Chat-GPT (0-shot) & 0.4447 & 0.5234 & 0.4642 \\

        \hline \hline
        \multicolumn{4}{|l|}{\textit{Our methods that leverage label descriptions}}\\
        \hline
        \classifier{} & 0.7125 & 0.7527 & 0.7053 \\
         \genclassifier{} & \textbf{0.8525} & \textbf{0.8966} & \textbf{0.8885} \\

    \hline
    \end{tabular}
    \caption{Performance of different models on the label classification task of the CAVES dataset. Our proposed model \genclassifier{} performs the best in all metrics. 
    }
    \label{tab:multi-label-results}
\end{table}


\vspace{2mm}
\noindent \textbf{Classification results}:
Table~\ref{tab:multi-label-results} states the performance of all models over the CAVES test set.
Our proposed generative variant \genclassifier, consistently demonstrates superior performance across all evaluation metrics whereas our discriminative variant \classifier{} becomes the second best. Among the baselines, CT-BERT performs the best.

Our generative classifier \genclassifier{} achieves 19.64\% higher macro-F1 score compared to our discriminative variant, \classifier. 
\genclassifier{} achieves an even more impressive 23.12\% increase in performance over the top-performing baseline model, CT-BERT.
The greater increase in Macro-F1 indicates that our model performs better on the niche classes that are present in smaller numbers in the dataset. 
Thus, incorporating class-descriptions into a model improves its ability to deal with such niche classes.

\begin{table*}[t]
\centering
\small
       \begin{tabular}{|p{79mm}|p{50mm}|p{16.4mm}|p{15mm}|}
        \hline
        \textbf{Tweet Text}  &\textbf{Generated Description} & \textbf{GT Labels} & \textbf{Pred labels}\\
        \hline
        \hline
        Pfizer is gonna make 15 BILLION in covid vax sales. Idc how your feeling now. I don’t understand ppl like you saying I’m fine right now. What if in a year you have cancer or an autoimmune illness. Both are possible.... & 
        Against Big Pharma — The tweet indicates that the Big Pharmaceutical companies are just trying to earn money, or is against such companies in
        & pharma, side-effect &  pharma
 \\
 \hline
        Mad fascist dictator Johnson now wants to force you to take an experimental vaccine with no long-term safety profile. All this for a virus with 0.3\% IFR. 1922 committee must step in NOW and REMOVE this communist lunatic! &  UnAgainst mandatory vaccination — The tweet suggests that vaccines should not be made mandatory.Untested / Rushed Process — The tweet
        & mandatory, unnecessary, rushed & rushed, mandatory
 \\
 \hline

        With the media pushing hard for vaccine passports/id. Boris has one chance left to prove that he is indeed, conservative. If he caves in and divides the people further, he will destroy what is left of his credibility. &  Against mandatory vaccination — The tweet suggests that vaccines should not be made mandatory.     Political Political Political Political & mandatory &  mandatory, political
\\
\hline
    \end{tabular}
    \caption{Examples of tweets where \genclassifier{} gives partially correct predictions -- either wrongly predicts only a subset of the ground truth (GT) labels or more than the GT labels.}
    \label{table:err_analysis}
\end{table*}

\vspace{2mm}
\noindent \textbf{Error analysis of best classifier}:
We performed an error analysis of the examples misclassified by \genclassifier{}. For a majority of the cases, we found the predictions by \genclassifier{} to be partially correct, such as predicting one out of two gold-standard labels correctly, or predicting a wrong label along with a correct label. Some examples of such partially correct predictions are given in Table~\ref{table:err_analysis}.

\section{Data Preparation for Analysis}
\label{sec:data}

This section describes how we collect data from Twitter, and prepare for a long-term analysis of the anti-vax discourse.

\subsection{Tweet Collection}

Our goal is to analyse the anti-vaccine discourse (specifically, anti-vaccine tweets ) ranging from pre-COVID to post-COVID times. 
To this end, we first used the Twitter Academic API~\footnote{
\url{https://developer.twitter.com/en/docs/twitter-api}} to collect vaccine-related tweets posted during the 5-year span between Jan. 2018 and Jan. 2023. 
We searched tweets using various keywords highlighted in \citet{poddar2022caves,poddar2022winds,gunaratne2019temporal}. The keywords included a combination of generic vaccine-related keywords (e.g. `vaccine', `vaxxer') and COVID-vaccine specific keywords (e.g. `astra zeneca', `pfizer', `comirnaty', `moderna'). 

Also, for analysing the impact of COVID-19 on \textit{non-COVID} vaccines, we collected tweets containing mentions of 4 common and highly-debated vaccines -- (i)~Flu Vaccine, (ii)~MMR (Measles, Mumps, Rubella) vaccine, (iii)~IPV (Inactivated Polio) vaccine and (iv)~HPV (Human Papillomavirus) / Gardasil vaccine. 
For this, we used the keywords such as `MMR', 'measles vaccine', `HPV', `gardasil', etc.

Using all of these keywords, we collected about 95M (million) distinct vaccine-related tweets (excluding retweets) posted in between January 2018 and January 2023. 

\new{
Though our data collection strategy does not guarantee the collection of all the vaccine-related tweets, we believe that the wide variety of keywords used to collect data ensures a fair representation of the overall vaccine discourse on Twitter. Since we perform analyses on aggregate data over time-periods of several months (as reported in subsequent sections), we believe the insights gained represent the overall broad picture of the vaccine discourse on Twitter.
}

\subsection{Identifying Anti-Vaccine Tweets}

In this work, since we are specifically interested in the anti-vaccine discourse,  we used the 3-class vaccine stance classifier developed in our prior work~\cite{poddar2022winds} on all the tweets we collected, to identify the anti-vaccine tweets. 
This 3-class CT-BERT-based classifier gives the probability of a vaccine-related tweet being Anti-vax or Pro-vax or Neutral, and achieves macro-F1 scores in the range [0.780, 0.858] on various datasets~\cite{poddar2022winds}.
We identified tweets that were predicted as \textit{Anti-Vax} by this classifier with probability $\geq 0.8$ -- the same threshold as was used for identifying anti-vax tweets for creation of the CAVES dataset~\cite{poddar2022caves}.
This high threshold is chosen to minimize chances of misclassification.
Through this process, we were finally left with \textbf{19.6M Anti-Vax tweets} in total.

\subsection{Classification of Pre-COVID / Non-COVID tweets}
\label{sub:precovid}

\begin{table}[tb]
    \centering
    \small
    \begin{tabular}{|l|c|c|c|}
    \hline
        \textbf{Model} & \textbf{Macro-F1} & \textbf{Weighted-F1} & \textbf{Jaccard} \\
        \hline
        \hline 
        \classifier{}  & 0.7389 & 0.7813  & 0.7125 \\
        \genclassifier{}  & \textbf{0.8215} & \textbf{0.8216}  & \textbf{0.7909} \\
   
    \hline
    \end{tabular}
    \caption{Performance of our classifiers on our \dataset{} dataset containing tweets about non-COVID vaccines. All models are trained on the CAVES training set.}
    \label{tab:novid-results}
\end{table}

As stated earlier, our classifiers were trained on the CAVES dataset which contains only COVID vaccine-related tweets posted during the pandemic.
But we want to apply our classifier on tweets spanning a long duration including both pre-COVID and COVID times, as well as on tweets related to non-COVID vaccines; hence we wanted to check if our classifier performs well on such tweets too. 

\new{To this end, we created a dataset of 500 randomly selected tweets (from among all the  anti-vax tweets we collected) containing (i)~anti-vax tweets posted during pre-COVID times,  and (ii)~anti-vax tweets posted during/after the COVID-19 pandemic, but about the four different \textit{non-COVID vaccines} (as mentioned earlier in this section).} 
We asked 3 undergraduate students (who are well-versed in English, frequent Twitter users, and none of whom is an author of this paper) to annotate the 500 tweets in an exactly similar way as the CAVES dataset. 
The class labels and descriptions (Table~\ref{tab:classes}) were informed to them, and they were asked to ``label each tweet with one or more appropriate labels \textit{independently}'' (without consulting each other).
The labels marked by at least 2 of them were retained while tweets with no majority labels were discarded, similar to what was done for the CAVES dataset~\cite{poddar2022caves}. 
Finally, this dataset (which we call \textbf{\dataset{}}) contains 486 tweets labelled into the same 12 labels as in the CAVES dataset. 

We applied our classifiers on this new dataset \dataset{}. Note that all these models were trained over the CAVES training data; we used the \dataset{} set only for testing.
The results over \dataset{} are given in Table~\ref{tab:novid-results}. Our models, particularly \genclassifier{}, perform well on this dataset as well, depicting its robustness. 
Thus we conclude that our classifiers can effectively be used on any anti-vaccine tweet, both related to COVID vaccines and non-COVID vaccines.

\subsection{Time Periods for analysing the anti-vax discourse}

We want to analyse people's anti-vaccine concerns during different points of time with respect to the COVID-19 pandemic. Accordingly, we consider these 4 time periods:

\noindent 1) \textbf{Pre-COVID period (Jan 2018 - Jan 2020)}: This period was before the COVID-19 outbreak was declared as a global health emergency by WHO (on 30 January 2020). 

\noindent 2) \textbf{COVID-start period (Feb 2020 - Dec 2020)}: The time of the pandemic, but before a vaccine for COVID-19 was made available to the general public. 

\noindent 3) \textbf{COVID-vax period (Jan 2021 - Apr 2022)}: The period when newly developed COVID-19 vaccines were being administered at large-scale.

\noindent 4) \textbf{Post-COVID period (May 2022 - Jan 2023)}: This is the period after COVID-19 cases/deaths started declining (as per WHO reports\footnote{\url{https://tinyurl.com/who-covid-update-may-2022}}) and vaccination rates stagnated as people stopped fearing the disease\footnote{\url{https://tinyurl.com/nyt-covid-vaccine-decline}}.

\section{Analysis}
\label{sec:analysis}

We now perform various analyses on the anti-vaccine tweets posted during the four time periods 
to understand how COVID-19 affected the anti-vaccine discourse. 
Note that we excluded the `religious' and `country' anti-vax classes from our analyses since they are substantially rarer than the other classes, having less than 1\% tweets each.

\subsection{RQ2: How has the Anti-Vax discourse changed due to COVID-19?}

\begin{figure}[t]
     \centering
     \begin{subfigure}[b]{0.884\linewidth}
         \centering
         \begin{tikzpicture}
        \node[anchor=south west,inner sep=0] (image) at (0,0) {\includegraphics[width=\linewidth,trim={4mm 5mm 3mm 3mm},clip]{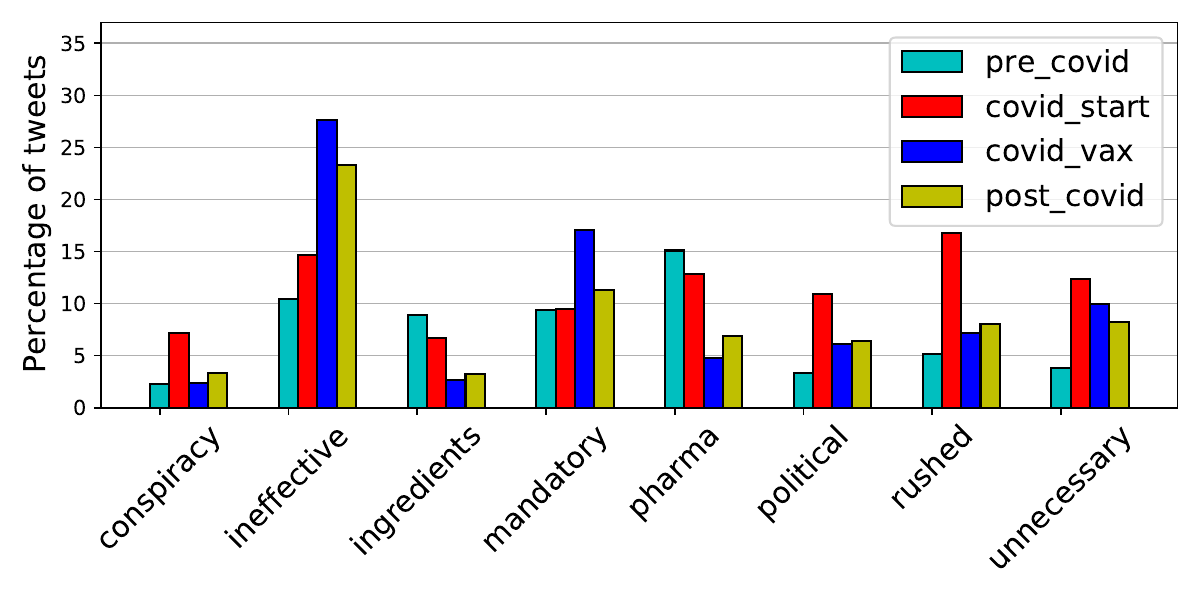}};
        \begin{scope}[x={(image.south east)},y={(image.north west)}]

            \draw[gray, thick] (0.28,0.65) -- (0.355,0.71);
            \draw[gray, thick] (0.28,0.65) -- (0.355,0.963);
            \node[anchor=south west,inner sep=0] (image) at (0.35,0.7) {\includegraphics[width=0.173\linewidth,trim={4mm 4.5mm 3mm 3mm},clip]{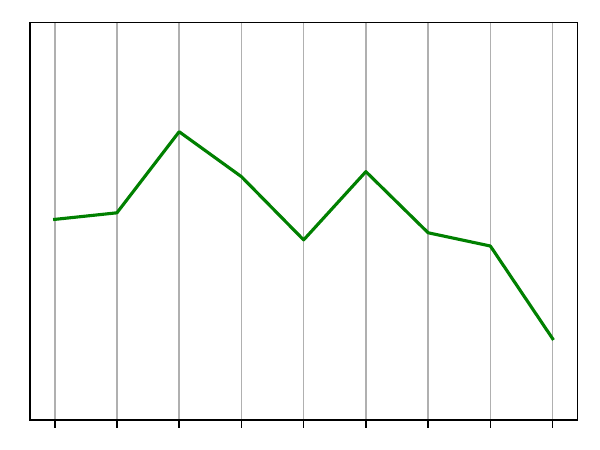}};
        \end{scope}
    \end{tikzpicture}
     \end{subfigure}
     \hfill
     \begin{subfigure}[b]{0.105\linewidth}
         \centering
         \includegraphics[width=\textwidth,trim={3mm 0mm 3mm 3mm},clip]{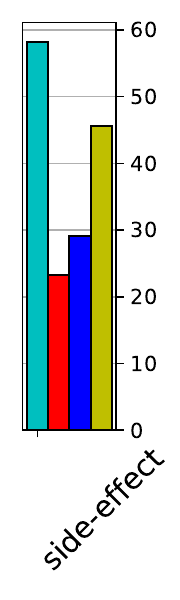}
     \end{subfigure}
        \caption{Distribution of concerns in anti-vax tweets across the different time-periods. The `side-effect' class (the largest class) is shown separately, to enable better visualization of the other classes. The inset graph shows the decline of the ineffective class in post-COVID period. }
        \label{fig:allvac_dist}
\end{figure}

To characterize the anti-vax discourse during a certain time-period, we apply our classifier over all anti-vax tweets posted during that period, and compute the distribution of the different concerns (labels given by the classifier).
Figure~\ref{fig:allvac_dist} compares this distribution of anti-vax concerns over all the four time-periods. 
Note that, since a tweet can have more than one concern stated in them, the sum of individual percentages for a time period may exceed 100\%. 
Also, we show the `side-effect' class separately (on the right), since it is by far the largest class, and including it in the same figure would make it difficult to visualize the temporal variations in the other classes (we consistently show the `side-effect' class separately in all subsequent figures).

\vspace{2mm}
\noindent \textbf{Dispersion of concerns during the COVID pandemic:}
From Figure~\ref{fig:allvac_dist}, we see that during pre-COVID times, the two primary concerns about vaccines were the `\textit{side-effect}' and `\textit{pharma}' classes. 
In the side-effects class, people mainly talked about Autism among children allegedly from the MMR vaccine, and the side-effects of the HPV vaccine in teenagers. 
The `pharma' class was the other major class with accusations against the big pharmaceutical companies of scams.
These two classes accounted for almost 75\% of the anti-vaccine concerns.

The focus on `side-effects' and `pharma' declined since the onset of the COVID-19 pandemic (the COVID-start period). Especially, the proportion of tweets discussing `side-effects' declined rapidly, as people started discussing other concerns. 
For instance, many people started arguing that vaccines for COVID-19 were \textit{`unnecessary'} (e.g., since COVID-19 had a low mortality rate) and/or `\textit{ineffective}'. 
The `rushed' class also gained prominence with netizens raising concerns regarding COVID vaccines not being tested well enough. 
The proportion of tweets posting `conspiracy theories' and `political' reasons also increased significantly during the COVID-start period. 
Thus, the \textit{anti-vax discourse became a lot more dispersed/varied during the COVID times} (as compared to the pre-COVID times), as several different anti-vax concerns started to be prominently discussed.

\vspace{2mm}
\noindent \textbf{Anti-vax discourse during the post-COVID times:}
Since May 2022 (the post-COVID period), we see some of the concern-classes returning towards their pre-COVID proportions; but there still remain some lingering differences from the pre-COVID times. 
For the classes `ineffective', `mandatory', `political' and `unnecessary',  we notice the fractions of tweets in the post-COVID period  decreasing (compared to their respective fractions during the COVID-vax period), thus bringing these classes closer to their proportions in the pre-COVID period.
To bring this out more clearly, 
the inset graph in Figure~\ref{fig:allvac_dist} shows the `ineffective' class decreasing gradually (fractions shown month-wise since May 2022).

However, there are still some major lingering differences from the pre-COVID distribution. 
For instance, a lot more tweets still express concerns with vaccines being `ineffective' and `mandatory' than in the pre-COVID times. 
In contrast, the proportion of tweets in the `pharma' class continues to be a lot lower than in the pre-COVID times. 
It remains to be seen whether these differences persist in the future, or if the anti-vax discourse returns to its pre-COVID state.

\subsection{RQ3: How has COVID-19 impacted the anti-vax discourse around non-COVID vaccines?}

We now concentrate on uncovering the changes in anti-vax discourse around non-COVID vaccines due to the pandemic. 
This analysis holds practical importance, since, these non-COVID vaccines are the main vaccines to be administered regularly even after the pandemic has subsided. 
Thus the changes in their discourse, if any, need to be identified.

\begin{figure}[!t]
     \centering
     \begin{subfigure}[b]{0.884\linewidth}
         \centering
         \includegraphics[width=\textwidth,trim={4mm 5mm 3mm 3mm},clip]{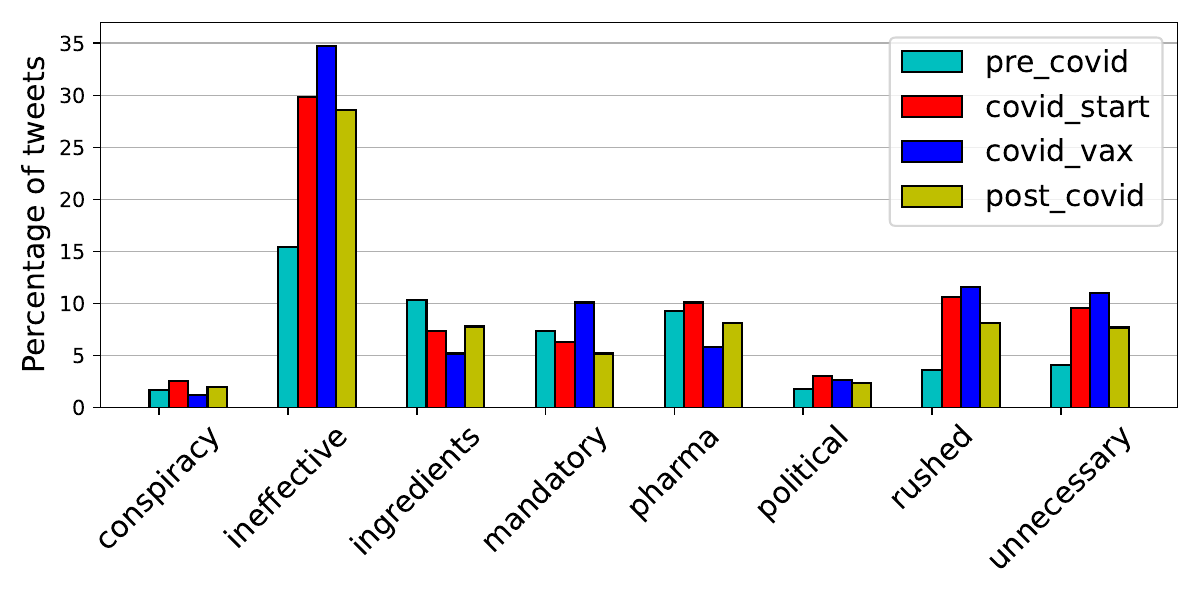}
     \end{subfigure}
     \hfill
     \begin{subfigure}[b]{0.105\linewidth}
         \centering
         \includegraphics[width=\textwidth,trim={3mm 0mm 3mm 3mm},clip]{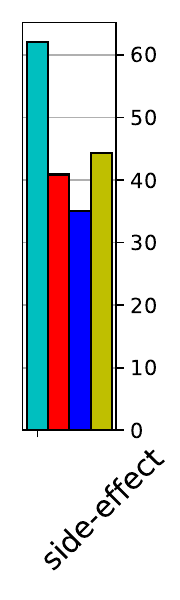}
     \end{subfigure}
        \caption{Distribution of concerns among anti-vax tweets about non-covid vaccines across the different time-periods. }
        \label{fig:novid-dist}
\end{figure}

To this end, we filter anti-vax tweets posted about four non-COVID vaccines that are heavily debated over the last several decades -- (i)~Flu vaccine, (ii)~MMR, (iii)~IPV and (iv)~HPV vaccine, \new{as discussed in the previous section.} 
Figure~\ref{fig:novid-dist} shows the distribution of concerns among all the anti-vax tweets about these vaccines, across the different time-periods. 
Note that the distributions in Figure~\ref{fig:novid-dist} are different from those in  Figure~\ref{fig:allvac_dist}, since Figure~\ref{fig:novid-dist} considers only those tweets that mentioned one of the above four non-COVID vaccines, while Figure~\ref{fig:allvac_dist} included all anti-vaccine tweets.

\vspace{2mm}
\noindent \textbf{Changes in the distribution of concerns:}
For some of the concern-classes, e.g. `conspiracy' and `political', we observe very low variations in the discourse around non-COVID vaccines (Figure~\ref{fig:novid-dist}), unlike what was observed for the overall distributions in Figure~\ref{fig:allvac_dist}. 
These concerns about the non-COVID vaccines seem to be mostly unaffected by the pandemic. However, the other concern-classes have been affected.  
For the `side-effect' and `ingredients' classes, we observe a slight dip in the percentages since the start of the pandemic, which reduced further during the COVID-vax period. 
In contrast, for the `ineffective', `rushed' and `unnecessary' classes, the fraction of tweets rose during the COVID times, and then slightly more during the COVID-vax period, but have gone down during the post-COVID period. 

If we compare the pre-COVID discourse and the post-COVID discourse about non-COVID vaccines, we find an increase in the `ineffective', `rushed' and `unnecessary' classes. In particular, there has been a huge rise in the `ineffective' concern-class. 
On the other hand, the `mandatory' concern has lessened during the post-COVID period (which is in contrast to what was observed for the overall distribution). This is mostly due to people talking less against mandating of the HPV vaccine, in the post-COVID times. 
Thus, some of the concerns around non-COVID vaccines have indeed been affected by the COVID-19 pandemic.

Till now, we have only analysed the temporal variations in the relative fractions of the concern-classes (Figure~\ref{fig:novid-dist}).
Furthermore, manual observation of the tweets reveals several interesting trends in the \textit{sub-topics being discussed about the non-COVID vaccines at different time-periods}.
We now describe these trends.

\begin{table}[t]
    \centering
    \small
    \begin{tabular}{|p{79.6mm}|}
    \hline
        So we know the flu vaccine causes other respiratory issues in kids. Now is the covid vaccine doing the same thing with kids and \#RSV? \\
        \hline
        Flu vaccine started in 1938; 80 years of Flu vaccine development and today it’s only 45\%-60\% effective.  Covid19 vaccine was developed in 8 months and we're told it's 95\% effective !\\         
        \hline
        Gates' vaccine found to cause Polio... so by all means let him vaccine the world for COVID?? \\
        \hline        
        So \#covid19 drug probably mandatory despite taking 12 months, not the usual 12 years. The last Mandatory one (MMR) saw a friends 2 healthy sons become very Autistic\\
    \hline
    \end{tabular}
    \caption{Examples of tweets (excerpts) posted during COVID-start/COVID-vax periods, where existing concerns about \textit{non-COVID} vaccines were projected to COVID vaccines. Such tweets often have a sarcastic tone.}
    \label{tab:novid-on-covid}
\end{table}

\vspace{2mm}
\noindent \textbf{Trends during COVID-start/COVID-vax times:}
We noticed that in many tweets posted during the COVID-start/COVID-vax periods, several \textit{existing concerns about the non-COVID vaccines were projected towards the COVID vaccines}.
Some examples of such tweets (often sarcastic) are given in Table~\ref{tab:novid-on-covid}.
For instance, we see concerns about the side-effects of Flu vaccine (respiratory issues) and the ineffectivenss of Flu vaccine being projected onto COVID vaccines (e.g., in the $1^{st}$ and $2^{nd}$ tweets in Table~\ref{tab:novid-on-covid} respectively).
Several Twitter users were also seen associating the role of Bill Gates in disseminating Polio Vaccines (IPV) with his role in developing COVID vaccines  (e.g., $3^{rd}$ tweet in Table~\ref{tab:novid-on-covid}).
Again, people expressed concerns that, similar to some non-COVID vaccines in the past, the development/testing of COVID vaccines were being too rushed. 
These observations partially explain some of the trends in Figure~\ref{fig:novid-dist}, such as the steep rise in the `ineffective' and `rushed' concerns during the COVID-start and COVID-vax periods.

\begin{figure}[t]
    \centering
    \includegraphics[width=0.83\linewidth,trim={4mm 5mm 3mm 3mm},clip]{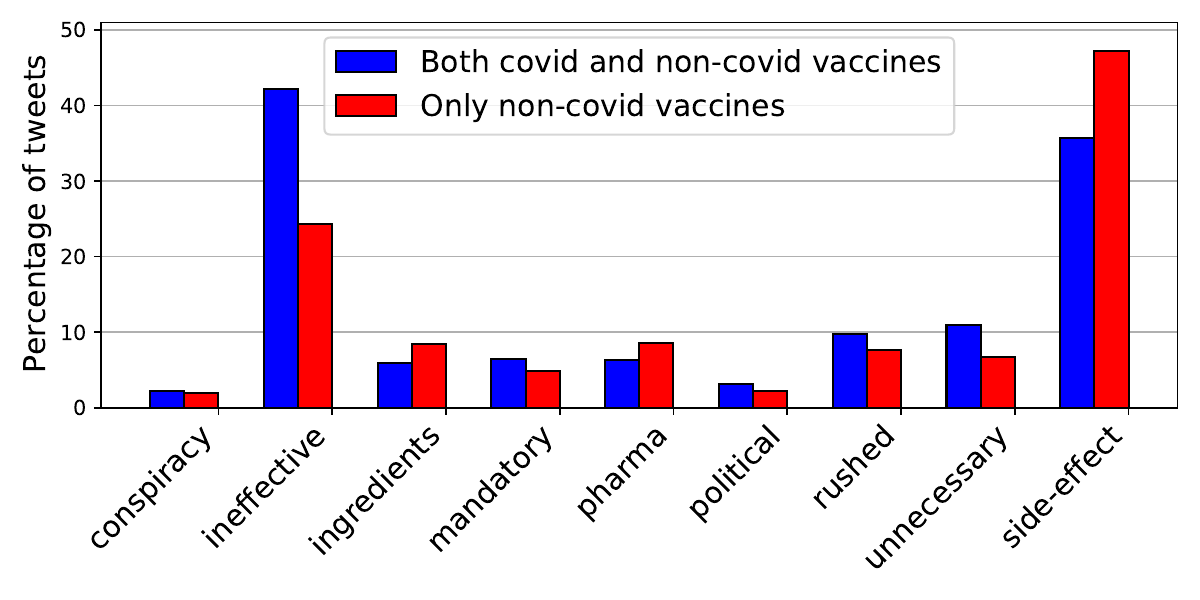}
    \caption{Distribution of concerns in tweets from post-COVID, (i)~that mention both COVID and non-COVID vaccines, and (ii)~that mention only non-COVID vaccines.}
    \label{fig:novid-post-dist}
\end{figure}

\vspace{2mm}
\noindent \textbf{Trends during post-COVID times:}
We observe that the discourse around non-COVID vaccines has become very varied and complicated in the post-COVID era.
In the post-COVID period, we observe two subsets of tweets mentioning the non-COVID vaccines -- (i)~about 25\% of the tweets  mention \textit{both non-COVID and COVID vaccines}, and 
(ii)~the rest 75\% tweets mentioning only non-COVID vaccines, and {\it not} COVID vaccines. 
The distributions of anti-vax concerns in these two types of tweets are shown in Figure~\ref{fig:novid-post-dist}.

\begin{table}[!t]
    \centering
    \small
    \begin{tabular}{|p{79.6mm}|}
    \hline
    \textbf{Type-I (23\%): Concerns about COVID vaccines, but not about non-COVID vaccines} \\
        \hline
        Inventor of polio vaccine offered it to the world for free, no profit for himself. Pfizer and others offer the Covide vaccine at a cost to humanity. That's the greedy corporation for you. Even the vaccine is not fully effective.\\
        \hline
        If you've been vaccinated for Chicken Pox, Polio, Small Pox you actually didn’t get these diseases. How about COVID ... haven't you or anyone you know got COVID after getting the vaccine? So maybe it's not really a vaccine.\\
        
    \hline \hline
    \textbf{Type-II (70\%): Covid has enhanced \textit{prior concerns} about non-COVID vaccines} \\
    \hline
        Now that the world knows that Covid19 vaccinations were created to kill and mame the masses, will we now speak about how the MMR Vaccine was created to cause Autism? \\
        \hline
        Flu vaccines don’t work for the same reason Covid injections don’t work – The Expose \\
     
    \hline
    \hline
    \textbf{Type-III (7\%): \textit{New concerns} that arose regarding COVID vaccines getting projected onto non-COVID vaccines}\\
    \hline
        The same experimental mRNA technology that is in the Covid19 jabs and boosters will also be in future Flu shots. The best bet is to say NO to all vaxxines \\
        \hline
        The Dangers of Adapting COVID19 Antigens into Childhood Vaccines -- Dr. Andrew Wakefield: ``It is going to be a disaster'' \\
        \hline
        I'm an RN, and I too will never get another flu shot. I've had 2 flu shots in my life. No more. And no covid vaccine for me \\  
        \hline
        I will not be getting a flu shot 4 the first time in many years. Never a covid booster! Not getting a pneumonia shot either. But I've lost all confidence in CDC protocols. They can't be trusted \\
    \hline
    \end{tabular}
    \caption{Examples of anti-vax tweets (excerpts) from the \textit{post-COVID} period, where both COVID-vaccines and Non-COVID vaccines have been mentioned. We see three distinct types of tweets (detailed in the text).}
    \label{tab:novid-post-covid}
\end{table}

The tweets that mention only non-COVID vaccines (and do not mention COVID vaccines) show high similarity with the pre-COVID tweets (when only non-COVID vaccines were discussed). The KL-divergence of the distribution of concerns in these tweets with respect to that in the pre-COVID tweets is low ($3.29$), as also evident by comparing the red bars in Figure~\ref{fig:novid-post-dist} with the pre-covid distribution in Figure~\ref{fig:novid-dist}. 
A manual observation of 100 randomly-sampled tweets that mention only non-COVID vaccines, also shows them to be very similar to the anti-vax tweets posted during pre-COVID times.
However, the other set of tweets that contain mentions about both non-COVID and COVID vaccines are very different. Their distribution of concerns shows high differences from the pre-covid distribution (KL-divergence of $38.71$). 
Hence it becomes more interesting to study these tweets since they contain lingering effects of the COVID-19 pandemic over the discourse arond non-COVID vaccines.

A manual observation of 200 tweets that were posted during post-COVID and mention both COVID and non-COVID vaccines, shows 3 broad types of tweets (examples of each type in Table~\ref{tab:novid-post-covid}). 
First, in about 23\% of the tweets, the users say that though they were concerned about COVID vaccines, they do \textit{not} have those concerns about non-COVID vaccines (Type-I, in the top part of Table~\ref{tab:novid-post-covid}).
Second, in about 70\% of the tweets, we see users who were unwilling to take any vaccine from pre-COVID times, and now their prior concerns about non-COVID vaccines have been enhanced in the post-COVID era (Type-II). Some examples of such tweets have been given in the middle part of Table~\ref{tab:novid-post-covid}.
Third, in the post-COVID era, we find signs of the opposite effect that we observed in the COVID-start/COVID-vax phases -- about 7\% of the tweets reflect that \textit{new concerns that arose about COVID vaccines are being projected onto the non-COVID vaccines} (Type-III). 
Some such examples are depicted in the bottom part of Table~\ref{tab:novid-post-covid}. One major talking point was about a new flu vaccine being developed with the controversial mRNA technology that the Pfizer and Moderna COVID vaccines were based on. Other examples include people being sceptical to take vaccines, unwilling to trust pharmaceutical companies after COVID-19.

\vspace{2mm}
\noindent \textbf{Emergence of a new set of anti-vax users}:
Note that, within Type-III, we observed some users who earlier readily took the non-COVID vaccines (during pre-COVID times), but are now sceptical of these vaccines after the pandemic (e.g., see the last two tweets in Table~\ref{tab:novid-post-covid}). 
Thus, it seems that, in addition to the \textbf{\textit{traditional anti-vaxxers}} who consistently post anti-vax sentiments both before and after the pandemic, a new set of users has emerged, who supported vaccines before the pandemic, but are unwilling to take vaccines now;
we call this new set of users as the \textbf{\textit{converted anti-vaxxers}}.
We study these users in detail in the next section.

\subsection{RQ4: What are the concerns of Traditional and Converted Anti-Vaxxers after the pandemic?}

\begin{figure}[t]
     \centering
     \begin{subfigure}[b]{\linewidth}
         \centering
         \includegraphics[width=0.855\linewidth,trim={4mm 5mm 3mm 3mm},clip]{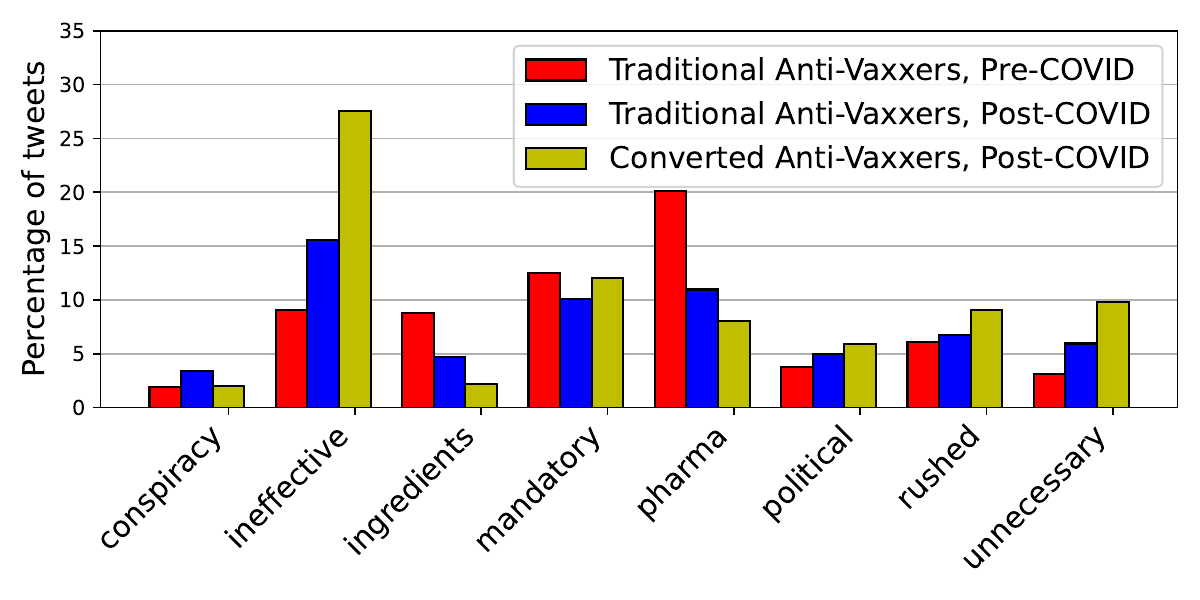}
         \includegraphics[width=0.1\linewidth,trim={3mm 0mm 3mm 3mm},clip]{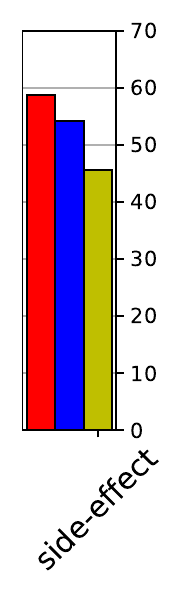}
         \caption{All vaccine tweets}
         \label{fig:user_allvac}
     \end{subfigure}
     \hfill
     \begin{subfigure}[b]{\linewidth}
         \centering
         \includegraphics[width=0.855\linewidth,trim={4mm 5mm 3mm 3mm},clip]{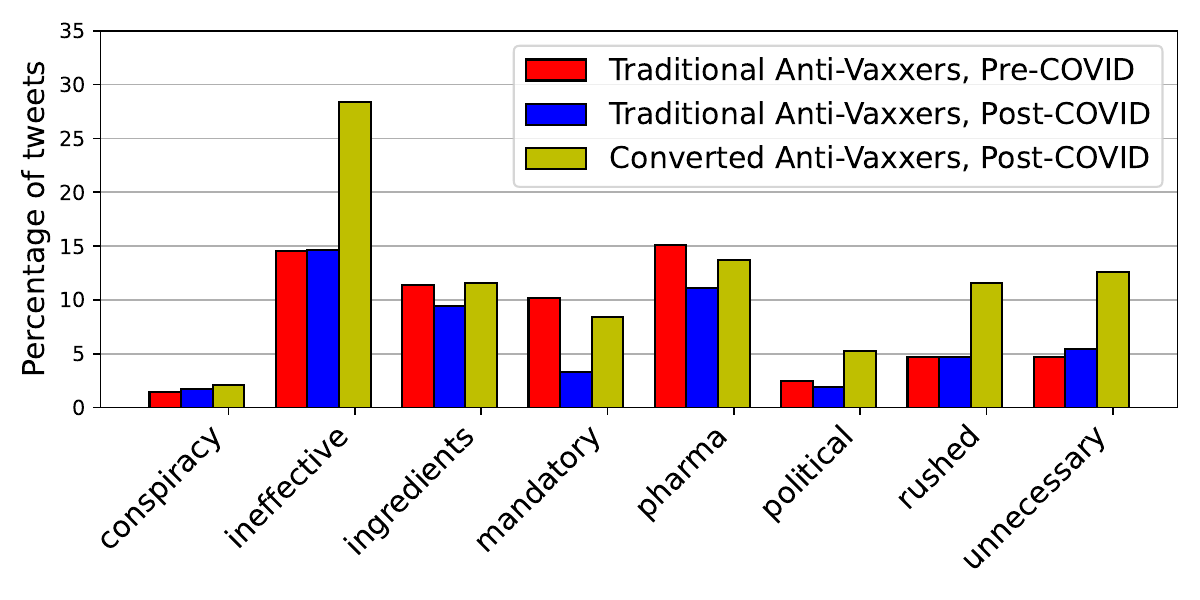}
          \includegraphics[width=0.1\linewidth,trim={3mm 0mm 3mm 3mm},clip]{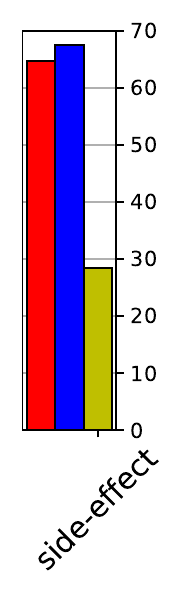}
         \caption{Non-COVID vaccine tweets}
         \label{fig:user_novid}
     \end{subfigure}
    \caption{Distribution of anti-vaccine concerns posted by the traditional anti-vaxxers during pre- and post-COVID periods (red and blue bars), and the converted anti-vaxxers during post-COVID (yellow bars).} 
    \label{fig:user_wise}
\end{figure}


A set of people used to post anti-vax opinions on Twitter since a long time~\cite{kata2012anti}, though this constituted only a small fraction of Twitter users. Many of these users continue to post anti-vax tweets during and after COVID times.
We call this group of users as the \textit{``Traditional Anti-Vaxxers''}.
After the onset of the COVID-19 pandemic, some users who were originally pro-vaxxers, also started showing hesitancy towards taking vaccines (as was observed in~\cite{poddar2022winds}). 
In the previous section, we also observed a set of people who took vaccines before COVID, but are unwilling to do so now (e.g. the users who posted the last two tweets in Table~\ref{tab:novid-post-covid});
we call this group of people the \textit{``Converted Anti-Vaxxers''}.
This observation motivated us to study these two groups of users, to understand their anti-vaccine concerns.

\vspace{2mm}
\noindent \textbf{Identifying Traditional and Converted Anti-vaxxers:}
To label a user as an Anti-Vaxxer (respectively, Pro-Vaxxer) during a particular time-period, we use the same method as suggested by \citet{poddar2022winds} -- we check if a user has posted at least $3$ vaccine-related tweets during that period, with $\geq \!70\%$ anti-vax (respectively, pro-vax) tweets among them.\footnote{Note that we are using a relatively strict criterion, to minimize the chances of misclassifying the vaccine-stance of users.}
Following this criterion, we got 13,776  anti-vaxxers and 52,403 pro-vaxxers during the pre-COVID times. 
Out of these anti-vaxxers, 1,388  are found to satisfy the same criterion during post-COVID times as well, i.e., they regularly posted anti-vax tweets both before and after the pandemic. We identify this set as the {\it traditional anti-vaxxers}. 
We also find 271 \textit{pre-COVID pro-vaxxers} who posted at least 3 vaccine-related tweets with $\geq \!70\%$ \textit{anti-vax} tweets during post-COVID; we identify this set as the \textit{converted anti-vaxxers}.
We also manually examined 50 randomly selected users from the two groups to check if they are automated bots; none of them seemed to be bots.

\vspace{2mm}
\noindent \textbf{Change in concerns of Traditional Anti-Vaxxers:}
Figure~\ref{fig:user_wise}(a and b) compares the distribution of concerns of the traditional anti-vaxxers during pre-COVID and post-COVID times.
If we look at the overall distribution considering all vaccine-related tweets (the red and blue bars in Figure~\ref{fig:user_allvac}), the traditional anti-vaxxers talk more about the `conspiracy', `ineffective' and `unnecessary' classes during post-COVID, compared to the pre-COVID times. 
In contrast, the `mandatory' and `side-effect' classes seem to have dipped slightly, while the `ingredients' and `pharma' classes have dropped a lot in the post-COVID times, compared to the pre-COVID times.
If we look at the distribution in those tweets that mention non-COVID vaccines (the red and blue bars in  Figure~\ref{fig:user_novid}), we observe that the post-COVID trends have mostly reverted to those in pre-COVID for most classes. 
However, there is a notable drop in the `mandatory' class, and slight drop in the `pharma' and `ingredients' classes. 

A manual study of 100 randomly selected tweets from this group of users shows almost all of them to be of Type-II described in the previous subsection (Table~\ref{tab:novid-post-covid}), where they continue to reject all vaccines and discuss how COVID has enhanced their prior concerns about the vaccines.

\vspace{2mm}
\noindent \textbf{Change in concerns of Converted Anti-Vaxxers:}
We show the distribution of the concerns of the converted anti-vaxxers during post-COVID in Figure~\ref{fig:user_wise}(a and b), and we compare it with that of the traditional anti-vaxxers during the same period (i.e., comparing the blue and yellow bars in Figure~\ref{fig:user_wise}).
We observe that the converted anti-vaxxers talk a lot more about the \textit{ineffectiveness} of both the COVID and non-COVID vaccines (compared to the traditional anti-vaxxers). 
The `rushed' and `unnecessary' classes are also discussed more by the converted anti-vaxxers, than the traditional anti-vaxxers. 
On the contrary, the `side-effect' class is less commonly discussed by the converted anti-vaxxers.

A manual analysis of 100 randomly selected tweets from this group of users reveals the following trends.
The converted anti-vaxxers mostly post about the COVID vaccines, which seem to be their primary concern even in post-COVID times.
When they post about the non-COVID vaccines, 
some of them speak about how COVID vaccines are less effective in comparison to non-COVID vaccines (type~I tweets in Table~\ref{tab:novid-post-covid}).
Others in this group post about the concerns about COVID vaccines being projected onto the non-COVID vaccines (type~III in Table~\ref{tab:novid-post-covid}).
They frequently express concerns about the mRNA technology being used in non-COVID vaccines (primarily the Flu vaccine) in post-COVID times.
Also they post against the pharmaceutical companies and the political influence over vaccines.
These concerns are making them unwilling to take the non-COVID vaccines, particularly the Flu vaccine, in post-COVID times.

\section{Conclusion}
\label{sec:conc}

In this work, we propose two novel classifiers that incorporate label descriptions to accurately identify the specific anti-vaccine concerns expressed in tweets. 
Then we apply the best classifier on anti-vaccine tweets posted between 2018 and 2023 to derive insights on how the vaccine discourse has been affected due to the pandemic.
We find that the concerns have become much more varied and complex since the onset of the pandemic. 
Alarmingly, many users have associated the concerns about COVID vaccines (e.g., use of mRNA) onto the non-COVID vaccines, and are thus hesitant to take vaccines such as the Flu vaccine in the post-COVID period. 

\vspace{2mm}
\noindent \textbf{Implications and broader impact:}
Our work provides tools to authorities to understand the specific concerns that people have towards vaccines.
Our proposed classifier (\genclassifier{}) is a key contribution that can be used by authorities to automatically gain insights on the specific anti-vax concerns of a person or a group of persons. 
Using this, authorities can take the following actions --

\noindent \textbf{(i)~Personalise counter-arguments:} 
We have seen that the vaccine concerns have become a lot more varied since the pandemic and generic counter-arguments may not be enough to reduce an individuals' vaccine-related concerns. 
For example, someone concerned about the side-effects needs a different counter-argument to someone who does not trust big pharma. 
Our approach will help authorities apply targeted/personalised counter-arguments depending on the specific concerns one has about vaccines.

\noindent \textbf{(ii)~Mitigate concerns towards non-COVID vaccines:}
We see that many people are associating their concerns about COVID vaccines with the non-COVID vaccines too, e.g., talking about the ineffectiveness of the Flu vaccines, side-effects of MMR and HPV, and so on. 
Many people are also outraged over the mRNA technology allegedly being incorporated in non-COVID vaccines.
Identifying such specific concerns about different vaccines can help authorities address individual concerns in a systematic way. 

\noindent \textbf{(iii)~Prevent conversion of users to anti-vaxxers:}
We observed that there are users who supported vaccines earlier (pre-COVID times) but are turning against them in post-COVID times (whom we studied as the `converted anti-vaxxers). 
Identifying these users and sensitizing them about the benefits of vaccines over their specific concerns can help restore their faith in vaccines.

\vspace{2mm}
\noindent \textbf{Limitations of our work}:
One of the limitations is that much of the vaccine-related data was scraped retrospectively, and many tweets (especially anti-vax tweets) could have been deleted. 
Also, our dataset depends on the list of keywords we used, and could have missed some tweets. 
However, given the large set of our collected data and the large and varied set of keywords we used, we hope that the high-level insights drawn in this study have not been affected too much, even if some tweets were missed.

Another limitation is that our data analysis depends on the classification of our classifiers, which can result in misclassification errors. However, note that we have taken strict thresholds (e.g. $0.8$ for identifying a tweet as anti-vax) to minimize the possibility of these errors.



\vspace{2mm}
\noindent \textbf{Ethical considerations:}
\new{
We have tried our best not to raise privacy concerns for the users whose Twitter posts we analyzed. 
We report all results over tweets aggregated over several months (and from several users for RQ4). The released data also has the usernames obfuscated.
It should be noted that different people can have varying degrees of concerns towards vaccines, and though they may not be traditional anti-vaxxers, they may still be hesitant towards vaccines in some particular aspects. In this work we use the term  `Anti-vaxxers' for all such users, just to follow the common terminology as prior works.  
Additionally, while labelling a user as Anti-Vax~(for RQ4), we follow very strict criteria -- we consider a tweet as anti-vax only if the classifier predicts the label with probability $\geq 0.8$, and we consider a user as anti-vax only if he/she posted $>70\%$ anti-vax (and at least $3$) tweets during a time period~(as also done in prior works~\cite{poddar2022winds}).  
Though this is not a foolproof method, we believe that outliers will not affect the insights much, given that we derive insights from data aggregated over several months and from several users.
}

\vspace{2mm}
\noindent \textbf{Future work:}
\new{There are several directions of future work. First, the classification performance of the models may be improved by using ensemble of various classifiers, or by using custom decoding strategies for the generative model.}
We also observed that the anti-vaccine discourse has become quite complex; for instance, nowadays there are many tweets that talk in support of the traditional vaccines but are against COVID vaccines. 
Thus, traditional sentiment analysis tools may not be sufficient for such tweets. 
Rather, application of Aspect Based Sentiment Analysis (ABSA) methods to such tweets can be an interesting future work.
Another future work could be to build automated methods for effectively countering anti-vaccine sentiments using personalised approaches.

\new{
Finally, we understand that getting large-scale data from the Twitter/X API has gotten very difficult and expensive, since early-2023. 
Nevertheless, we believe this work lays the groundwork for future analysis of vaccine-related concerns from other types of text as well. Our classifiers and analysis process can easily be modified to run on short-message texts from other social media websites. 
The classifiers can also be made to run on longer posts from websites like Reddit by using models like Longformer and Long-T5.
}

\vspace{4mm}
\noindent \textbf{Acknowledgements:} We thank the anonymous reviewers for their comments which helped to improve the paper. We also acknowledge the students who annotated the tweets for this study. The first author (S. Poddar) is supported by the Prime Minister’s Research Fellowship (PMRF) from the Ministry of Education, Government of India.

{
\small
\bibliography{ref}
}


\end{document}